\documentclass[sigconf]{acmart}
\def\BibTeX{{\rm B\kern-.05em{\sc i\kern-.025em b}\kern-.08emT\kern-.1667em\lower.7ex\hbox{E}\kern-.125emX}}

\copyrightyear{2020}
\acmYear{2020}
\setcopyright{acmlicensed}
\acmConference[CIKM]{The Conference on Information and Knowledge Management}{2020}{Online}
\setcopyright{none}
\renewcommand\footnotetextcopyrightpermission[1]{}
\usepackage{pifont}
\usepackage{multirow}
\newcommand{\cmark}{\ding{51}}%
\newcommand{\xmark}{\ding{55}}%

\usepackage{subcaption}
\begin{document}
\sloppy
%
\title{DeText: A Deep Text Ranking Framework with BERT}

\author{Weiwei Guo, Xiaowei Liu, Sida Wang, Huiji Gao, Ananth Sankar, Zimeng Yang, Qi Guo, Liang Zhang, Bo Long, Bee-Chung Chen and Deepak Agarwal}
\affiliation{
  \institution{LinkedIn, Mountain View, California}
}
\email{{wguo, xwli, sidwang, hgao, ansankar, zyang, qguo, lizhang, blong, bchen, dagarwal}@linkedin.com}

%
\renewcommand{\shortauthors}{Guo et al.}

\begin{abstract}
Ranking is the most important component in a search system. Most search systems deal with large amounts of natural language data, hence an effective ranking system requires a deep understanding of text semantics. Recently, deep learning based natural language processing (deep NLP) models have generated promising results on ranking systems. BERT is one of the most successful models that learn contextual embedding, which has been applied to capture complex query-document relations for search ranking. However, this is generally done by exhaustively interacting each query word with each document word, which is inefficient for online serving in search product systems. In this paper, we investigate how to build an efficient BERT-based ranking model for industry use cases. The solution is further extended to a general ranking framework, DeText, that is open sourced and can be applied to various ranking productions. Offline and online experiments of DeText on three real-world search systems present significant improvement over state-of-the-art approaches.



\end{abstract}


%
\maketitle

\section{Introduction}
Search systems provide relevant documents to users who are looking for specific information through queries. A user receives a list of ranked documents ordered by search relevance, where ranking plays a crucial role to model such relevance that directly affects consequential user interactions and experience. Most search systems deal with a large amount of natural language data from queries, profiles, and documents. An effective search system requires a deep understanding of the context and semantics behind natural language data to power ranking relevance.


Traditional ranking approaches largely rely on word/phrase exact matching features, which has a limited ability to capture contextual and deep semantic information. In the recent decade, deep learning based natural language processing technologies present an unprecedented opportunity to understand the deep semantics of natural language data through embedding representation~\cite{Huang:13}. Moreover, to enhance contextual modeling, contextual embedding such as BERT~\cite{devlin2019} has been proposed and extensively evaluated on various NLP tasks with significant improvements over existing techniques.

However, promoting the power of BERT in ranking is a non-trivial task. The current effective approaches~\cite{dai2019,nogueira2019,qiao2019} integrate BERT as an embedding generation component in the ranking model, with the input a concatenated string of query and document texts. BERT is then fine tuned with ranking loss. The inherent transformer layer~\cite{vaswani:17} in BERT allows direct context sharing between query words and document words, exploiting the power of contextual modeling in BERT to the greatest extent, as the query word embeddings can incorporate many matching signals in documents. This approach, in the category of interaction based models~\cite{Guo:16,xiong2017,Dai:18}, comes with a significant challenge in online serving: a) the heavy BERT computation on the fly is not affordable in a real world search system; and b) the interaction based structure, as applied to concatenated query and document, precludes any embedding pre-computing that can reduce computation.
To enable an efficient BERT-based ranking model for industry use cases, we propose to use representation based structure~\cite{Huang:13,Shen:14}. Instead of applying BERT to a concatenated string of query and document texts, it generates query and document embeddings independently. It then computes the matching signals based on the query and document embeddings. This approach makes it feasible for pre-computing document embedding; thus, the online system only needs to do BERT real-time computation for queries. By independently computing query and document embeddings, however, we may lose the enhancement on the direct context sharing between queries and documents at word-level~\cite{qiao2019}. This trade-off makes it a challenge to develop a BERT-based ranking model that is both effective and efficient.

In this work, we investigated the BERT-based ranking model solution with representation-based structure, and conducted comprehensive offline and online experiments on real-world search products. Furthermore, we extended the model solution into a general ranking framework, DeText (Deep Text Ranking Framework), that is able to support several state-of-the-art deep NLP components in addition to BERT. The framework comes with great flexibility to adapt to various industry use cases. For example, BERT can be applied for ranking components that have rich natural language paraphrasing; CNN can be applied when ease of deployment is a top concern for a specific system.

Beyond the ranking framework, we also summarized experience on developing an \textbf{effective and efficient} ranking solution with deep NLP technology, and how to balance effectiveness and efficiency for industry usage in general. We shared practical lessons of improving relevance performance while maintaining a low latency, as well as general guidance in deploying deep ranking models into search production.

The contribution of this paper is summarized below:
\begin{itemize}
    \item We developed a representation based ranking solution powered by BERT and successfully launched it to LinkedIn's commercial search engines.
    \item We extended the ranking solution into a general ranking framework, DeText, that can be applied to different search products with great flexibility. The code is open sourced for public usage.\footnote{\url{www.github.com/linkedin/detext}}
    \item We provided practical solutions and lessons on developing and deploying neural ranker models with deep NLP w.r.t. balance between efficiency and effectiveness.
    
    


\end{itemize}

\section{Related Work}
In this section, we first introduce how Deep NLP models extract text embeddings, discuss their application in ranking, and then introduce the status of ranking model productionization.

\subsection{Deep NLP based Ranking Models}
\label{sec:related-work-two-category}
There are two categories of deep NLP based ranking models: representation based and interaction based models.  \textit{Representation based} models learn independent embeddings for the query and the document. DSSM \cite{Huang:13} averages the word embeddings as the query/document embeddings. Following this work, CLSM/LSTM-RNN \cite{Shen:14,Palangi:16} encodes word order information using CNN\cite{Lecun:95}/LSTM\cite{Hochreiter:1997}, respectively. All these three works assume that there is only one field on the document side, and the document score is the cosine similarity score of the query/document embedding. NRM-F \cite{Zamani2018} adds more fields in the document side and achieves better performance.  One major weakness of representation based networks is the failure to capture local lexical matching, since the text embedding, \textit{e.g.}, a 128 dimensional vector, cannot summarize all the information in the original text.

To overcome the issue, \textit{interaction based models} compare each part of the query with each part of the document. In DRMM \cite{Guo:16}, a cosine similarity is computed for each word embedding in the query and each word embedding in the document. The final document score is computed based on the pairwise word similarity score histogram. K-NRM \cite{xiong2017} and Conv-KNRM \cite{Dai:18} extended DRMM by kernel pooling and pairwise ngram similarity, respectively. Recently, BERT \cite{devlin2019} has shown superior performance \cite{dai2019,nogueira2019,qiao2019} in ranking. It is considered an interaction based model, since the query string and document string are concatenated as one sentence, where transformer layer \cite{vaswani:17} compares every word pair in that sentence.

In experiments of previous works, interaction based methods usually produce better relevance results than representation based methods, at the cost of longer computation time introduced by the pairwise word comparison.  

\subsection{Productionizing Deep Neural Ranker}
Commercial search engines have a strict requirement on the serving latency. Despite better relevance performance, the interaction based ranking approaches are not scalable due to the heavy interaction computation. Therefore, to our best knowledge, the representation based approaches are generally used for production. 

With representation based approaches, existing work uses embedding pre-computing, either for documents~\cite{ramanath2018, yin2016} or for member profiles (personalization)~\cite{Grbovic2018}. It requires a huge amount of hard disk space to store the embedding, as well as a sophisticated system design to refresh the embeddings when there are any document/profile changes.

\begin{figure}
\begin{subfigure}{0.15\textwidth}
\includegraphics[width=\linewidth]{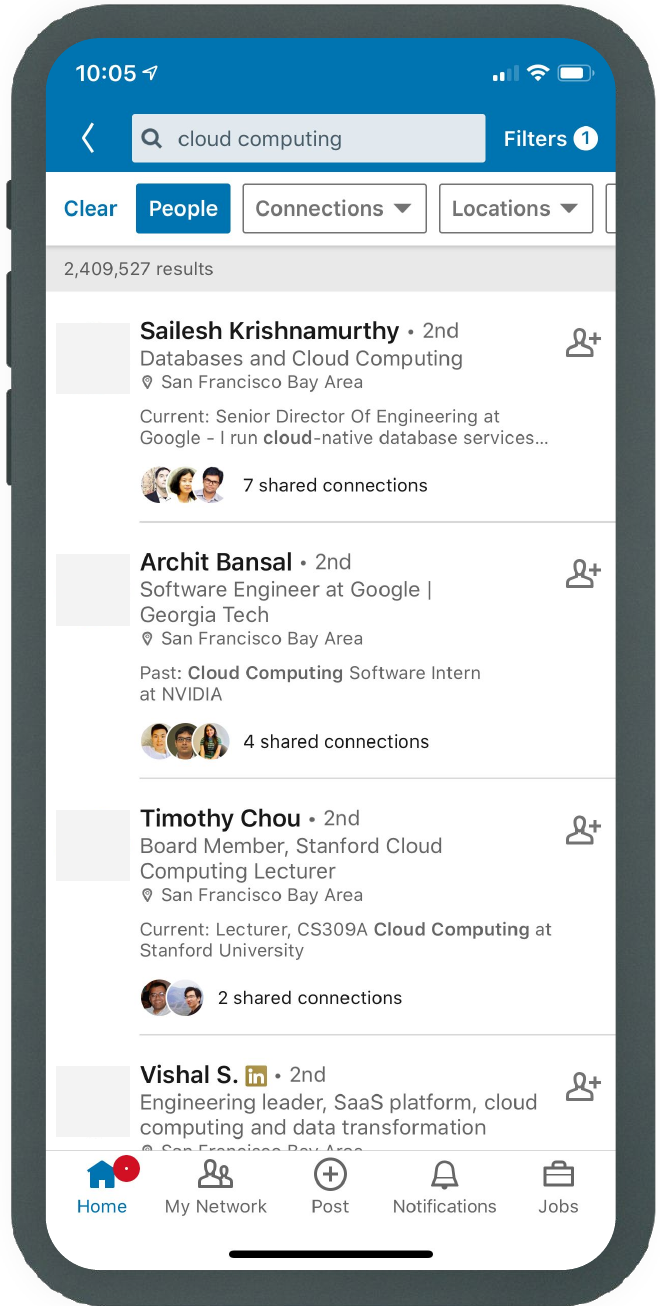}
\caption{People Search} \label{figure:people-search}
\end{subfigure}
\begin{subfigure}{0.15\textwidth}
\includegraphics[width=\linewidth]{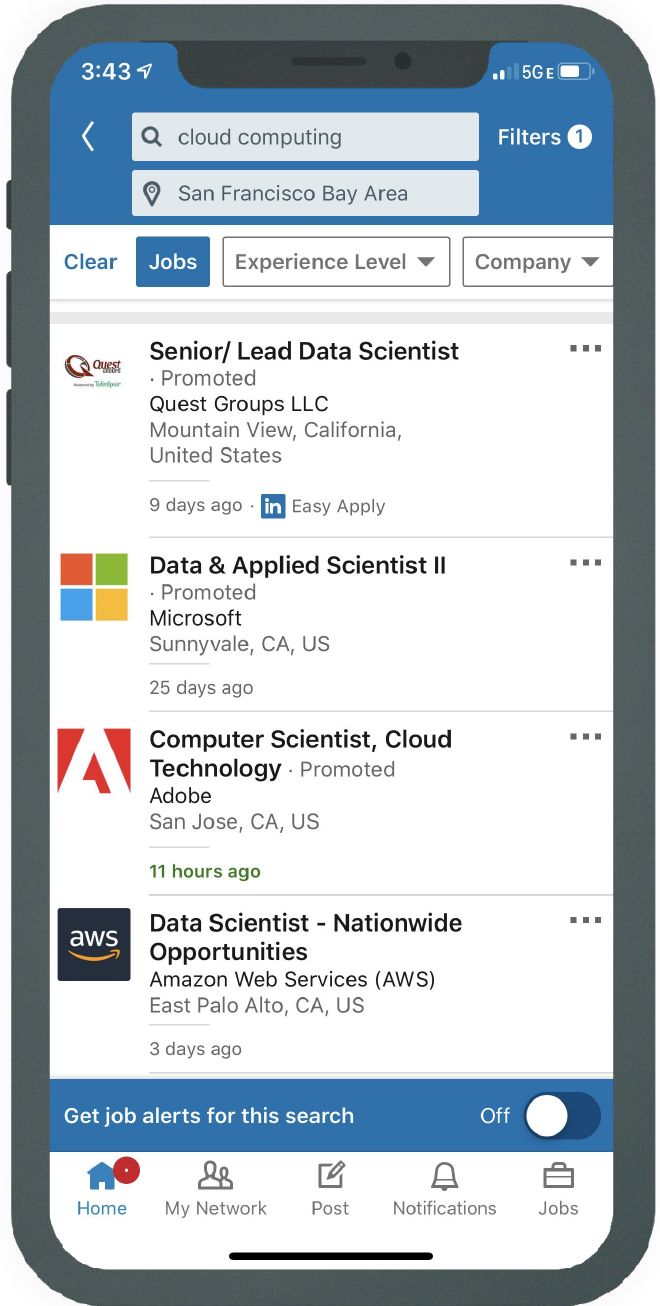}
\caption{Job Search} \label{figure:job-search}
\end{subfigure}
\begin{subfigure}{0.15\textwidth}
\includegraphics[width=\linewidth]{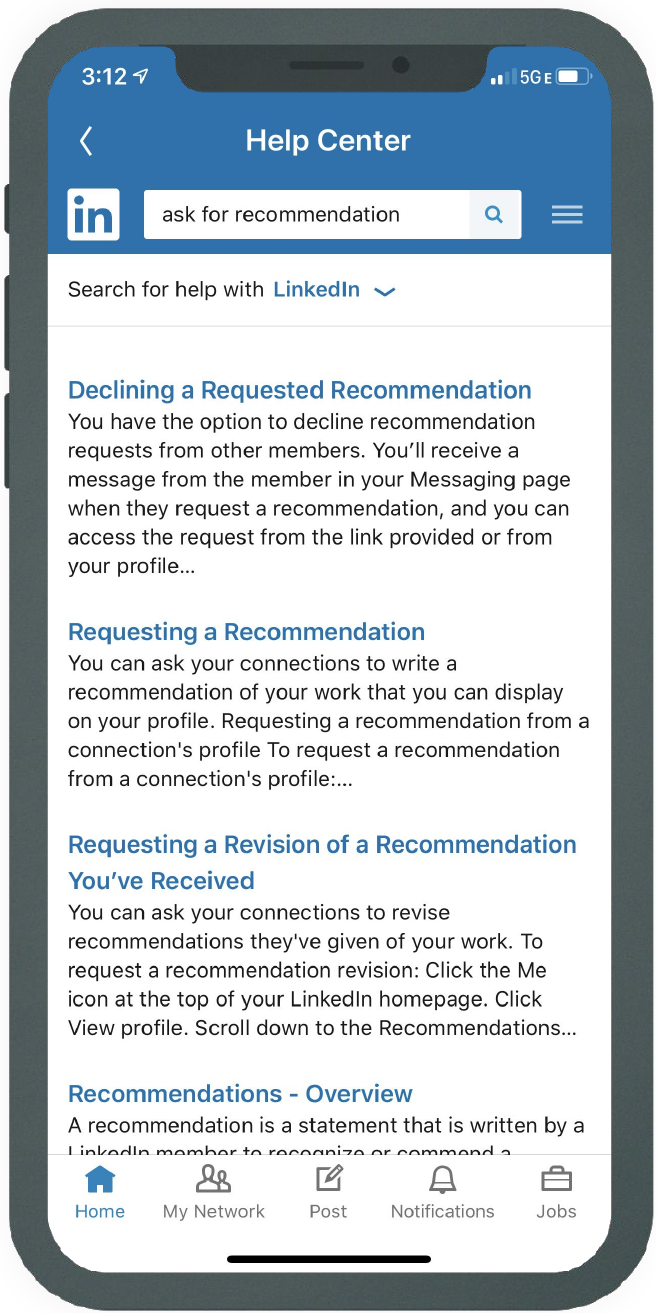}
\caption{Help Center} \label{figure:hc-example}
\end{subfigure}
\hspace*{\fill} 
\caption{The first two figures show the search result of "cloud computing" in people search/job search, respectively.  The last figure shows an example of query "ask for recommendation" in help center search.} \label{figure:search-example}
\vspace{-2mm}
\end{figure}

\section{Search Systems at LinkedIn}
\label{sec:linkedin-system}

\begin{table}
\small
  \caption{\small Summary of three vertical searches.}
  \vspace{-1mm}
  \label{table:data-set-stats}
  \begin{tabular}{lccc}
    \toprule
     & \textbf{People} & \textbf{Job} & \textbf{Help Center} \\
    \midrule
    \textbf{No. of Unique Docs} & 600M & 20M & 2,700 \\
  \bottomrule
\end{tabular}
\vspace{-3mm}
\end{table}

There are many search ranking systems at LinkedIn. Figure \ref{figure:search-example} shows three examples: people search that retrieves member profile documents; job search that ranks job post documents; and help center search that returns FAQ documents. The number of unique documents in each search system is listed in Table \ref{table:data-set-stats}. In general, the common part of these ranking systems is to discover the relevant documents, based on many hand-crafted features. Similar to other vertical searches such as Yelp or IMDB, the documents at LinkedIn are semi-structured with multiple fields. For example, member profiles contain headline, job title, company, etc. In general, the retrieval and ranking process needs to be finished around one or several hundred milliseconds.

The data from these three search verticals are different in nature.  
The queries and documents in help center search are the most similar to natural language, \textit{i.e.}, the text data is more likely to be a normal sentence with proper syntax, and majority queries are paraphrases of the problem that users want to address in help center search. People search is on the other end of the spectrum: the queries and documents are mostly entities without grammar; exact keywords matching such as company names is important. Job search data lies in between.

\section{DeText Framework for BERT-based Ranking Model}
In this section, we propose a BERT-based ranking framework using representation-based structure. The framework can be extended to support other neural network components, such as CNN and LSTM, for deep natural language processing. Specifically, we refer to the BERT-based ranking model as DeText-BERT, and directly illustrate the model using the open sourced DeText framework as shown in Figure~\ref{fig:detext-model}.

We design the DeText framework to be (1) general and flexible enough to cover most use cases of ranking modules in search systems; (2) able to reach a good balance between efficiency and effectiveness for practical use cases.

\vspace{-4mm}
\subsection{Architecture}
\label{section:framework}
\begin{figure*}
  \includegraphics[width=140mm]{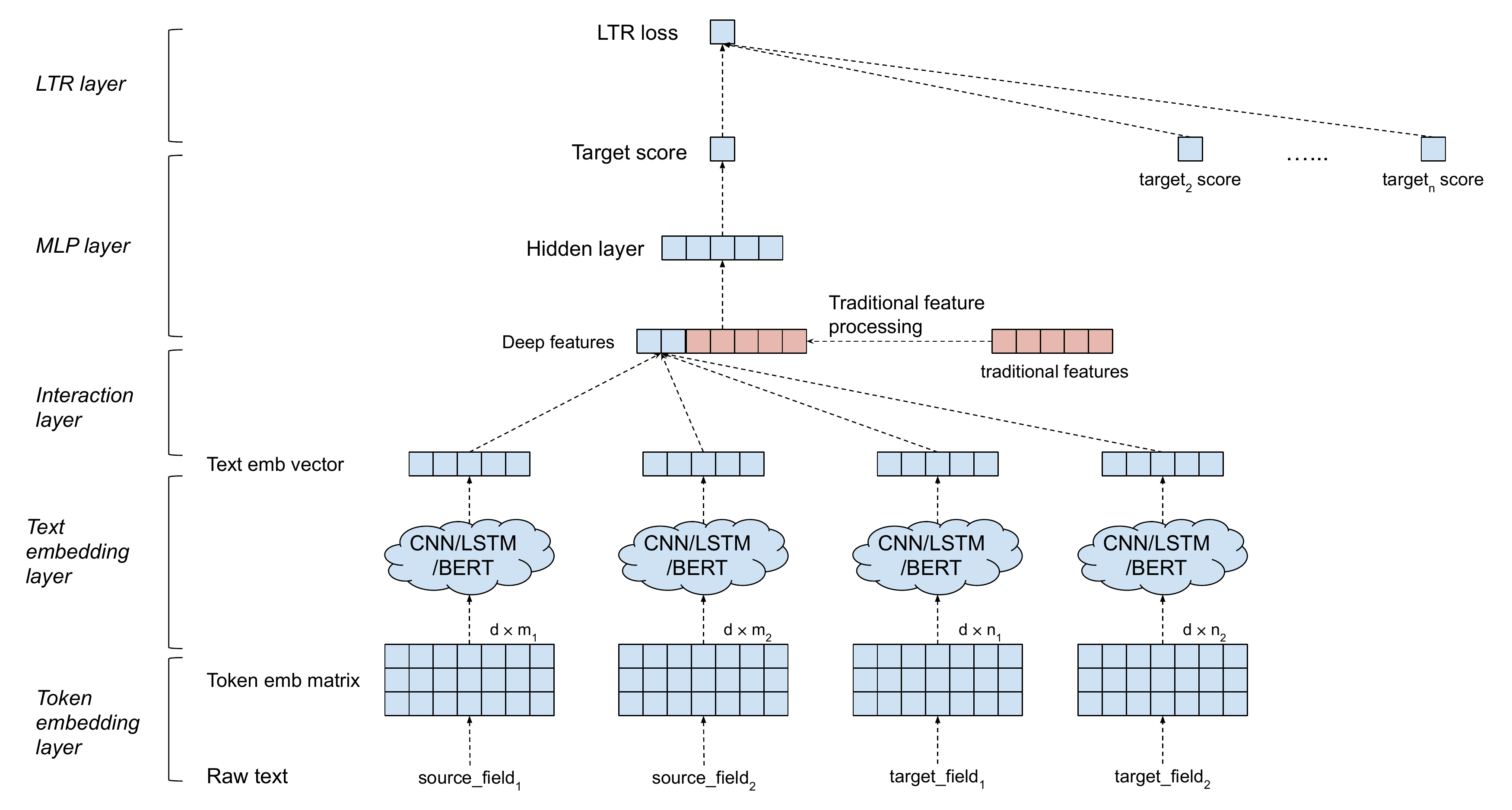}
  \caption{ The DeText framework.  In this figure, there are two source fields, and two target fields.}
  \label{fig:detext-model}
  \vspace{-3mm}
\end{figure*}

As illustrated in Figure \ref{fig:detext-model}, given multiple source (queries, user profiles) / target (documents) texts and traditional features, the DeText framework contains 6 components: Token Embedding Layer, Text Embedding Layer, Interaction Layer, Traditional Feature Processing, Multilayer-Perceptron Layer, and Learning-to-rank Layer. Specifically, DeText-BERT model uses BERT as the text embedding layer. In the rest of this section, we will illustrate the details of each component.

\textbf{Input Text Data}. The input text data is generalized as \textit{source} and \textit{target} texts. The source texts could be queries or user profiles. The target text could be documents. Both source and target could have multiple fields, which is different from most previous work \cite{Huang:13,Shen:14,Palangi:16}, where only two fields (query and document) are available. There are several advantages of using multiple fields: 
1). enable personalization with text fields from user profiles, and 2). achieve better and more robust results.

\textbf{Token Embedding Layer}. The sequence of text tokens is transformed into an embedding matrix $E$. For text with $m$ tokens, the matrix has a size of $d \times m$, where $d$ is the number of token embedding dimensions. Depending on the text encoding methods, different token granularities are used: in CNN/LSTM, the tokens are words; in BERT, the tokens are subwords \cite{sennrich2016}.

\textbf{Text Embedding Layer}. Under the representation based model structure, embedding is extracted independently for each text field. The embedding can be generated through various neural network components for deep natural language processing, such as BERT, CNN, LSTM, etc. The outcome of this layer is a $d$-dimensional embedding vector. More details are discussed in Section \ref{sec:bert-model} and \ref{sec:cnn-model}. 

\textbf{Interaction Layer}. The interaction between source and target only happens after the text embedding is generated, which is the key difference of representation based methods from interaction based methods.  Table \ref{table:interaction-features} summarizes the different interaction methods, where $u_s$/$u_t$ is the source/target field embedding, respectively. Note that for every source and target pair, cosine similarity generates one feature, while the Hadamard product/concatenation generates many features (a vector).
\begin{table}[h]
\small
\caption{Interaction features.}
\label{table:interaction-features}
\begin{tabular}{lcp{3.8cm}}
  \toprule
    Cosine similarity & $\frac{u_s^{\top} u_t}{\|u_s\|\cdot \|u_t\|}$ & one feature per source/target pair\\ [2mm]
    Hadamard product & $u_q \cdot u_d$ & $d$ features per source/target pair \\ [1mm]
    Concatenation & $u_q \oplus u_d$ & $d$ features per text field\\
  \bottomrule
\end{tabular}
\end{table}

\textbf{Traditional Feature Processing}.  
The existing hand-crafted features, such as personalization features, social networks features, user behavior features, are usually informative for ranking. To integrate them with deep NLP features, we use standard normalization and elementwise rescaling \cite{aksoy2001} to better process the features:
\begin{align*}
x_i^{(1)} &= \frac{x_i - \mu}{\sigma}\\
x_i^{(2)} &= wx_i^{(1)} + b
\end{align*}
where mean $\mu$ and standard deviation $\sigma$ are pre-computed from training data, and $w$ and $b$ are learned in the DeText-BERT model. 

\textbf{MLP Layer}. Deep features, as the output of the interaction layer, are concatenated with the traditional features as the final features, followed by a Multilayer-Perceptron (MLP) \cite{pal1992} layer to compute the final document score. The hidden layer in MLP is able to extract the non-linear correlations of deep features and traditional features.

\textbf{LTR Layer}. The last layer is the learning-to-rank layer that takes multiple target scores as input. DeText provides the flexibility of pointwise, pairwise or listwise LTR \cite{burges2010}, as well as Lambda rank \cite{burges2007}. Binary classification loss (pointwise learning-to-rank) can be used for systems where click probability is important to model, while pairwise/listwise LTR can be used for systems where only relative position matters.

\subsection{Flexibility of DeText}
The DeText framework enables model \textbf{flexibility} to adapt to demands of different productions, in terms of input data layer (multiple source/target fields), text embedding layer (CNN vs BERT), interaction layer (cosine/hadamard/concat), LTR (pointwise/pairwise/listwise), etc.

By enhancing the model flexibility, we can optimize the model \textbf{effectiveness} while maintaining \textbf{efficiency}. Firstly, representation based methods are used to bound the time complexity. Secondly, the flexibility of input data/interaction layer, together with traditional feature handling, enable us to experiment and develop scalable neural network models with strong relevance performance.

\subsection{DeText-BERT for Ranking}
\label{sec:bert-model}
To use BERT in ranking model, we follow the approach of fine-tuning on pretrained BERT model \cite{devlin2019}: The BERT model is firstly pretrained on unsupervised data, and then fine-tuned in ranking framework with supervised clickthrough data. To extract the text embedding, we use the embedding of a special token "[CLS]". The source/target embedding will later go through the interaction layer to generate deep features.

Previous work \cite{qiao2019} shows that directly training a representation based BERT ranking model does not yield good results. This is because the BERT fine-tuning requires a small learning rate (around 1e-5). Therefore, two optimizers are used in DeText, each with a different learning rate responsible for a different part of the model. For example, in our experiments (Section \ref{section:experiment-detext-model-setup}), we set 1e-5 for BERT components, and 1e-3 for other components. Using this dual learning rates strategy, a successful representation based ranking model with BERT can be trained. 

In order to reduce the online serving latency and capture domain-specific semantics, we also pretrained a compact BERT model on LinkedIn's in-domain data, named as LiBERT. More detailed can be found in Section \ref{section:bert-pretraining}.

\subsection{Support CNN for Ranking in DeText}
DeText framework can support CNN for deep natural language processing in the text embedding layer. It is worth noting that we use word tokens instead of triletters as in prior work~\cite{Huang:13,Shen:14}, since the latter lifts the computation (by an order of the character length of words). We follow the previous work \cite{kim2014} to generate the text embedding from word embedding matrix $E$. Specifically, it uses a one-dimensional CNN along the sentence length dimension. 
After max-pooling, the resulting text embedding vector has $f$ elements, where $f$ is the number of filters.

\section{Online Deployment Strategy}
\begin{figure}
  \includegraphics[width=82mm]{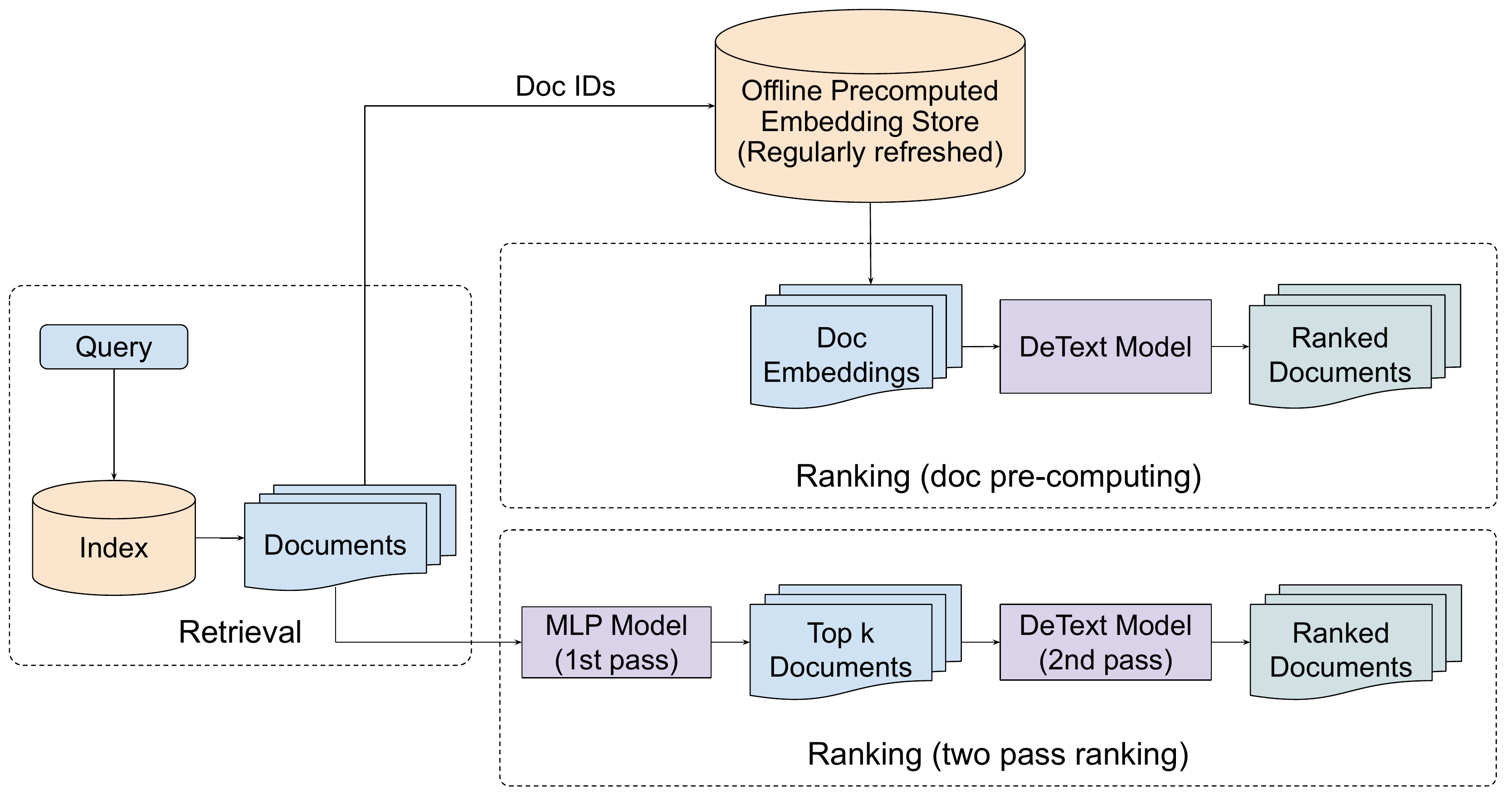}
  \caption{Document embedding pre-computing and two-pass ranking.}
  \label{figure:online-rank-2-modes}
\vspace{-3mm}
\end{figure}

The major challenge of deploying deep neural models comes from serving latency. As shown in Figure~\ref{figure:online-rank-2-modes}, two different deployment strategies, document pre-computing and two pass ranking, are designed for BERT based models and CNN based models, respectively. They are discussed in detail in the following subsections. Note that the online deployment strategies only affect ranking components; the document retrieval components stay the same.


\subsection{DeText-BERT with Document Embedding Pre-computing}
The multiple transformer layers in BERT is computationally time-consuming. Since DeText uses the representation based method, we are able to adopt the document embedding pre-computing approach for search ranking, as shown in the boxed section (doc pre-computing) in Figure \ref{figure:online-rank-2-modes}. For offline, document embeddings are pre-computed with BERT and saved in an embedding store, which is a fast key-value store where key is the document id, and value is the pre-computed document embedding vectors. The store is refreshed on a regular basis (e.g., daily). For online, after document candidates are retrieved from search index, the corresponding document ids are used to fetch the document embeddings from pre-computed embedding store. The benefit of this approach is to have the heavy BERT online computation only happen on the queries. It can significantly save online computation time, since the document texts are much larger than the query texts. In the setting of 10 documents for one query, the online latency can be reduced from hundreds of milliseconds to tens of milliseconds with this deployment strategy.

\subsection{DeText-CNN with Real-time Inference}
\label{sec:cnn-model}
DeText-CNN can be adopted with a different online integration strategy: real-time inference for both sources and targets.  Compared to document embedding pre-computing, real-time inference can simplify online system design without the need of pre-computing or refreshing document embeddings. Hence the real-time inference could be a lightweight solution readily applied for many search engines. 
In this paper, we show that real-time inference can be achieved by (1) choosing a compact DeText structure without hurting relevance performance too much, and (2) two pass ranking that reduces 99 percentile (P99) latency.


We find that a compact CNN structure with small dimensions can perform well in our experiments (Table \ref{table:num-cnn-filter}). This is mainly because traditional handcrafted features from the production systems already contain valuable information for ranking, so that the CNN model can be focusing on the signals that are missing in the traditional features.

Even with a simple network, the CNN computation time grows linearly with the number of retrieved documents. Therefore, we use a two pass ranking schema to bound the latency (Figure \ref{figure:online-rank-2-modes}, two pass ranking box). The first ranker is a MLP \cite{pal1992} with one hidden layer without the deep features, which is fast. After ranking, only the top ranked hundreds of documents are sent to the DeText-CNN model.\footnote{Note that the top hundreds of documents are in one worker, while the online ranking is distributed to many workers. Each worker is responsible for retrieving and ranking the documents on its own shard.}  This two pass ranking framework has several benefits: 1). easy to implement and deploy; 2). the MLP ranker can filter out a large amount of irrelevant documents, which provides a relatively small candidate set with high recall for CNN ranker; and 3). the latency is bounded since CNN is applied to a small set of candidates.

\section{Experiments}
In this section, we discuss the offline and online experiments of DeText on search ranking tasks in English traffic.

\begin{table}
\footnotesize
\caption{Online metrics definitions. 
}
\vspace{-2mm}
\label{table:online-metrics}
\begin{tabular}{lp{6cm}}
\toprule
    \textbf{Metric}&  \textbf{Definition} \\
\midrule
CTR@5 & Proportion of searches that received a click at top 5 items \\
Session Success Rate &  Proportion of search sessions that received a click. A new session is created if the user has no activity in help center search for 30 minutes.\\
Job Apply & Job search metric. Number of job applications from search. \\
Happy Path Rate & Help center search metric. Proportion of users who searched and clicked a document without using help center search again in that day, nor creating a ticket.\\
\bottomrule
\end{tabular}
\vspace{-2mm}
\end{table}

\subsection{Experiment Setting}

\subsubsection{Datasets}
\label{section:datasets}
The models are tested on three document ranking datasets, \textit{i.e.}, people search, job search, and help center search.  
The training data is collected from clickthrough data, which are sampled from 2 month traffic: 5 million queries for people search, 1.5 million queries for job search, 340 thousand queries for help center. Both development and test set have 50 thousand queries for each vertical from the later month. One query usually has 10 or more documents.  Multiple document fields are used as the target fields of DeText: 1). In people search, the documents are the member profiles; three profile fields are used: headline, current position, past position. 2). In job search, the job post title, company name are used.  3). In help center search, document title and example question (illustrates the typical question for the document) are used.


\subsubsection{Metrics} For both offline/online metrics, only relative metric improvement over baseline models instead of absolute values are presented, due to the company confidential policy. The online metrics are defined in Table \ref{table:online-metrics}.

\subsubsection{Baseline Models}
The production models are trained with XGBoost \cite{chen2016}. The hyper-parameters (pairwise vs listwise, number of trees, etc) are optimized by both manual tuning and auto hyper-parameter tuning, which are proven effective in LinkedIn's commercial search engines.

For each vertical search, there are existing hand-crafted traditional features. These features contain valuable information for the specific search engine verified by both offline and online experiments. The features can be categorized into three classes: 1). Text matching features. It includes not only exact matching features such as cosine similarity and jaccard similarity, but also semantic matching features, {\it i.e.}, named entity ids that are obtained by applying in-house entity linking tools \cite{shacham2017}; 2). Personalization features. For example, in people search, the social network distance between the searcher and the retrieved profiles; in job search, the searcher's title overlapping with the job post title; and 3). Document popularity features. For example, in people search, static rank of a member profile; in job search/help center, the clickthrough rate of a job post/FAQ document, respectively.


\subsubsection{DeText Models}\label{section:experiment-detext-model-setup} Two models are evaluated in this section: DeText-LiBERT (BERT model pretrained on LinkedIn data) and Detext-CNN. The default setting of DeText training is introduced below, unless specified otherwise: 1). \textit{Token embedding layer}: For DeText-CNN models, we always pretrain word embedding on the LinkedIn textual training data with Glove \cite{pennington2014}, which leads to comparable or better results than no word pretraining or existing word embedding trained on out-domain data. For DeText-LiBERT models, the word embeddings are from a BERT model pretrained on LinkedIn data. 2). \textit{Text embedding layer}: For DeText-CNN, the CNN filter window size is fixed as 3 for all text fields (we do not observe significant gain from using multiple window sizes), and the number of filters is fixed as 64.  For DeText-LiBERT, the model structure is described in Section \ref{section:bert-pretraining}. 3). \textit{Interaction layer}: The best combination of "cosine similarity and hadamard" is used for each dataset. 4). \textit{Feature processing layer}: Both normalization and element-wise re-scaling are performed on the traditional features. 5). \textit{MLP layer}: one hidden layer of size $200$. 6). \textit{Learning-to-rank layer}: we stick to listwise, since we find listwise ranking performs better (people search and job search) or comparable (help center) to pairwise ranking.

Regarding training, both DeText-CNN and DeText-LiBERT models are trained for 2 epochs. Adam optimizer \cite{kingma2015} is used with learning rate 1e-3; for the BERT component, the learning rate is 1e-5.  Each minibatch contains $256$ queries with associated documents.

\subsubsection{BERT Pretraining on LinkedIn Data}
\label{section:bert-pretraining}
The LinkedIn text data has many domain-specific terms, such as "Pinterest", "LinkedIn", resulting in a very different vocabulary from Wikipedia, as used by Google BERT (e.g., BERT$_{\text{BASE}}$) pretraining. Thus, we pretrained a LiBERT on domain specific data, and then fine tuned the parameters during DeText-LiBERT training. 

In order to reduce model serving latency, we use a smaller architecture compared to Google's BERT base model \cite{devlin2019}: (6 layers, 512 hidden, 8 heads). The resulting model has 34 million parameters, 1/3 of Google's BERT$_{\text{BASE}}$ model. We also use less data (around 1/5) than BERT$_{\text{BASE}}$ model, 685 million words vs 3.3 billion words. The statistics of LinkedIn data is listed in Table \ref{table:bert-data}. Although LiBERT pretraining is conducted in an unsupervised learning manner, we collected pretraining data from the time period prior to the three verticals' training data collection period, to ensure there is no potential data leaking bias. 

\begin{table}
\footnotesize
  \caption{LinkedIn data for BERT pretraining.}
  \vspace{-1mm}
  \label{table:bert-data}
  \begin{tabular}{lp{4cm}l}
    \toprule
     \textbf{Data Source} & \textbf{Description} & \textbf{\# Words}\\
    \midrule
    Search queries & Query reformulation pairs & 204M \\
    \multirow[t]{2}{*}{Member profiles} & \multicolumn{1}{l}{Member headlines and summaries} & \multicolumn{1}{l}{98M}\\
    & \multicolumn{1}{l}{Member position titles and descriptions} & \multicolumn{1}{l}{105M}\\
    Job posts & Job titles and descriptions & 217M \\
    Help center & Queries and doc titles & 61M \\
  \bottomrule
\end{tabular}
\vspace{-1mm}
\end{table}

\subsection{Search Ranking Experiments}
\begin{table}
\footnotesize
  \caption{\small Offline NDCG@10 score percentage lift in three searches over the production baseline XGBoost. $\dag$ and $\ddag$ denote statistically significant improvements at p < 0.05 using a two-tailed t-test over XGBoost and DeText-CNN, respectively.}
  \vspace{-1mm}
  \label{table:offline-ndcg}
  \begin{tabular}{lccc}
    \toprule
     \textbf{Models} & \textbf{People Search} & \textbf{Job Search} & \textbf{Help Center}\\
    \midrule
    DeText-MLP & $-0.07\%$ & $+0.05\%$ & $+0.15\%$ \\
    DeText-CNN &  $+3.02\%^{\dag}$ & $+4.65\%^{\dag}$ & $+11.56\%^{\dag}$\\
    DeText-LiBERT & $+3.38\%^{\dag\ddag}$ & $+6.14\%^{\dag\ddag}$ & $+13.94\%^{\dag\ddag}$\\
  \bottomrule
\end{tabular}
\vspace{-1mm}
\end{table}

\subsubsection{Offline Experiments}
All the relative percentage lift is calculated w.r.t the production baseline model.

\noindent\textbf{Overall: } Table \ref{table:offline-ndcg} summarizes the offline NDCG percentage lift in the three search datasets. To understand the impact of deep NLP on text data, we included one baseline model DeText-MLP (DeText with only MLP and LTR layers on traditional features). Since DeText-MLP does not use any text embedding, it has comparable results as XGBoost, which is also observed in previous works \cite{li2019}. For DeText-CNN, it consistently outperforms the strong production baseline model by a large margin. DeText-LiBERT is able to further improve the NDCG scores. The performance of DeText-CNN/DeText-LiBERT shows that deep learning models are able to capture a lot of semantic textual matching, hence a necessary  complement to the existing hand-crafted features.  

Meanwhile, it is worth noting that in Table \ref{table:offline-ndcg} deep learning models achieve the largest improvement on help center, followed by job search and people search. This is mainly caused by the genre of the data, as discussed in Section~\ref{sec:linkedin-system}: 1). In the help center, there are many paraphrases of the same scenarios, for example, query "how to hide my profile updates" to FAQ document "Sharing profile changes with your network". 2). In people search, exact matching is much more important as compared to the other two searches, for example, if the query contains the company word "twitter", generally we should not return a member profile who works at "facebook", even though the word embedding of the two companies could be similar. 3). Job search has less paraphrasing than help center, but more search exploration compared to people search.

\begin{table}
\footnotesize
  \caption{\small General BERT vs in-domain BERT on NDCG@10.}
  \vspace{-1mm}
  \label{table:linkedin-bert}
  \begin{tabular}{lccc}
    \toprule
     \textbf{Model} & \textbf{People Search} & \textbf{Job Search} & \textbf{Help Center}\\
    \midrule
    DeText-CNN &  $+3.02\%$ & $+4.65\%$ & $+11.56\%$\\
    \midrule
    DeText-BERT$_{\text{BASE}}$ & $+3.08\%$ & $+3.60\%$ & $+13.80\%$ \\
    DeText-LiBERT & $+3.38\%$ & $+6.14\%$ & $+13.94\%$ \\
  \bottomrule
\end{tabular}
\vspace{-1mm}
\end{table}

\noindent\textbf{LiBERT v BERT$_{\text{BASE}}$:} The impact of pretrained BERT on LinkedIn data is evaluated and shown in Table \ref{table:linkedin-bert}. In people search and job search, DeText-LiBERT significantly outperforms google's BERT$_{\text{BASE}}$, i.e., DeText-BERT$_{\text{BASE}}$, which should be attributed to the pretraining on in-domain data. In the help center where vocabulary and language are closer to Wikipedia, LiBERT can still achieve comparable results. It is worth noting LiBERT has only 1/3 of the parameters of BERT$_{\text{BASE}}$.

\begin{table}
\footnotesize
  \caption{\small Number of CNN filters in DeText-CNN on NDCG@10.}
  \vspace{-1mm}
  \label{table:num-cnn-filter}
  \begin{tabular}{lcccc}
    \toprule
     \textbf{\#Filters} &  \textbf{People Search} & \textbf{Job Search} & \textbf{Help Center}\\
    \midrule
    64 & $+3.02\%$ & $+4.65\%$ & $+11.56\%$\\
    128 & $+3.07\%$ & $+4.81\%$ & $+11.94\%$  \\
    256 & $+3.10\%$ &  $+4.82\%$ & $+12.37\%$ \\
    512 & $+3.16\%$ & $+4.92\%$ & $+12.74\%$ \\
  \bottomrule
\end{tabular}
\vspace{-1mm}
\end{table}

\begin{table}
\small
  \caption{\small Text embedding interaction in DeText-CNN on NDCG@10.}
  \vspace{-1mm}
  \label{table:text-emb}
  \begin{tabular}{lccc}
    \toprule
     \textbf{Interaction} & \textbf{People} & \textbf{Job} & \textbf{Help Center}\\
    \midrule
    cosine & $+2.67\%$ & $+4.25\%$ & $+11.02\%$\\
    cosine, hadamard & $\textbf{+3.02\%}$ & $+4.65\%$ & $\textbf{+11.56\%}$ \\
    cosine, concat & $+2.39\%$ & $+4.62\%$ & $+10.09\%$ \\
    cosine, hadamard, concat & $+2.84\%$ & $\textbf{+4.73\%}$ & $+11.49\%$ \\
  \bottomrule
\end{tabular}
\vspace{-1mm}
\end{table}

\noindent\textbf{Limit of CNN:} To better understand the trade-off on DeText-CNN models w.r.t. efficiency and effectiveness, we experimented with different numbers of CNN filters, shown in Table \ref{table:num-cnn-filter}. We observed with a large number of filters, the gain on people and job search is relatively small (less than $+0.4\%$). This is probably because the powerful hand-crafted features on people/job search are already integrated in the DeText model. Based on the results, we decided to adopt the CNN model with 64 filters in production to reduce the online serving latency.

\noindent\textbf{Text Embedding Interaction:} Table \ref{table:text-emb} shows the impact of different text embedding interaction methods. We used cosine similarity as a baseline, and gradually added features computed by other interaction methods. The experiments show that using both cosine similarity and hadamard product features can produce the best or 2nd best results. 

\begin{table}
\footnotesize
  \caption{\small Traditional features for DeText-CNN models on NDCG@10.  The first row does not use any traditional features.}
  \vspace{-1mm}
  \label{table:traditional-features}
  \begin{tabular}{cccccc}
    \toprule
    \textbf{Trad-ftr} &\textbf{Rescale} & \textbf{Norm} & \textbf{People} & \textbf{Job} & \textbf{Help Center}\\
    \midrule
    \xmark & \xmark & \xmark & $-4.52\%$ & $-9.98\%$ & $+11.07\%$ \\
    \cmark & \xmark & \xmark & $+2.31\%$ & $+3.17\%$ & $+11.13\%$ \\
    \cmark & \xmark & \cmark & $+2.47\%$ & $+3.44\%$ & $+11.55\%$ \\
    \cmark & \cmark & \xmark & $+2.71\%$ & $+4.49\%$ & $+11.24\%$ \\
    \cmark & \cmark & \cmark & $+3.02\%$ & $+4.65\%$ & $+11.56\%$\\
  \bottomrule
\end{tabular}
\vspace{-1mm}
\end{table}

\begin{table}
\footnotesize
  \caption{\small The impact of using multiple fields. In the single target field setting, the most important field is used: headline for people search and job post title for job search. In this experiment, all the traditional features are excluded. Note that DeText-CNN with a single target field is a special version of CLSM model \cite{Shen:14} that operates on words.}
  \vspace{-1mm}
  \label{table:one-doc-field}
  \begin{tabular}{lccc}
    \toprule
    \textbf{Model} & \textbf{\#fields} & \textbf{People} & \textbf{Job}\\
    \midrule
    DeText-CNN \small{(CLSM on words)} & single & $-5.20\%$ & $-12.83\%$ \\
    DeText-LiBERT & single & $-3.14\%$ & $-10.30\%$ \\
    \midrule
    DeText-CNN & multiple & $-4.52\%$ & $-9.98\%$ \\
    DeText-LiBERT & multiple & $-2.51\%$ & $-7.20\%$ \\
  \bottomrule
\end{tabular}
\vspace{-1mm}
\end{table}

\noindent\textbf{Traditional Features:} We evaluated the importance of processing traditional features, as shown in table \ref{table:traditional-features}. The first row, where no traditional features are used, proves that the traditional features are crucial in people/job search to capture social networks and personalization signals. In addition, both feature element-wise rescaling and normalization techniques are helpful; applying them together yields the best results.

\noindent\textbf{Multiple Fields:} Table \ref{table:one-doc-field} shows the impact of multiple document fields (using all the fields described in Section \ref{section:datasets}). To provide a dedicated comparison, we excluded the traditional features in this experiment. The results demonstrate that using multiple document fields can significantly improve the relevance performance. This is a practical solution for many real-world applications, since the documents in vertical search engines could be semi-structured with many text fields containing additional valuable information.

\subsubsection{Online Experiments}

\begin{table}
\footnotesize
  \caption{\small Online experiments of DeText-CNN and DeText-LiBERT.}
  \vspace{-2mm}
  \label{table:online-exp}
  \begin{tabular}{llll}
    \toprule
    \textbf{Search} & \textbf{Model} & \textbf{Metrics} & \textbf{Percentage Lift} \\
    \midrule
    People search & DeText-CNN & CTR@5 & $+1.13\%$\\
    & & Session Success Rate & neutral \\
    & DeText-LiBERT & CTR@5 & $+1.56\%$ \\
    & & Session Success Rate & $+0.23\%$ \\
    \midrule
    Job search & DeText-CNN & CTR@5 & $+3.16\%$\\
    & & Job Apply & $+0.73\%$ \\
    \midrule
    Help center & DeText-CNN & Happy Path Rate & $+15.0\%$ \\
    & & Session Success Rate & $+6.1\%$\\
    & DeText-LiBERT & Happy Path Rate & $+26.1\%$ \\
    & & Session Success Rate & $+11.1\%$\\
    \bottomrule
\end{tabular}
\vspace{-2mm}
\end{table}

We performed online experiments in the production environment with model candidates showing promising offline performance. The experiments are conducted with each model under at least 20\% traffic for more than two weeks, and the best models are later fully ramped to production. All reported metrics are statistically significant (p-value < 0.05) over the production baseline XGBoost. 


For LiBERT models, document embeddings are refreshed daily. However, in job search, there are many new job postings on an hourly basis, which requires the embedding precomputing in a more frequent manner such as near-line update. Due to the computational resources and product priority, we leave the online experiment of DeText-LiBERT on job search to future work.

Table \ref{table:online-exp} summarizes the experiments of DeText-CNN/DeText-LiBERT on three search engines. From CTR@5 on people and job search, we observed a similar trend in online/offline metrics: the improvement on job search is larger than on people search. Furthermore, DeText-LiBERT is consistently better than DeText-CNN in people search and help center, indicating the importance of contextual embedding on capturing deep semantics between queries and documents in search.


\subsubsection{Latency Performance}
To better understand the latency performance, the offline P99 latency on people search is provided in Table \ref{table:people-latency}. Similar patterns on job search and help center are observed, and they are not presented due to limited space. 
For each worker, there are thousands of documents to score. The CNN model in the two pass ranking will score hundreds of documents. All numbers are computed by a Intel(R) Xeon(R) 8-core CPU E5-2620 v4 @ 2.10GHz machine and 64-GB memory.

We also compared with another variant, all-decoding, that is to score all the retrieved documents on the fly. By comparing the first two settings in Table \ref{table:people-latency}, it proves two pass ranking is effective at reducing the P99 latency.  Meanwhile, the online A/B test does not show significant relevance difference between all-decoding and two pass ranking strategies. 

With the document precomputing strategy, we are able to fully ramp the DeText-LiBERT models to production within latency requirements. In addition, we are interested in the LiBERT performance w.r.t. BERT$_{\text{BASE}}$. Our experiments suggest that  DeText-LiBERT is faster than DeText-BERT$_{\text{BASE}}$, due to the smaller model structure of the former.

\begin{table}
\footnotesize
  \caption{\small The latency at 99 percentile on people search.}
  \vspace{-2mm}
  \label{table:people-latency}
  \begin{tabular}{llll}
    \toprule
    \textbf{Model} &\textbf{Deployment Strategy} & \textbf{Time}\\
    \midrule
    DeText-CNN People & all-decoding & +55ms \\
    DeText-CNN People & two pass ranking & +21ms\\
    DeText-LiBERT people & doc pre-computing & +43ms \\
    DeText-BERT$_{\text{BASE}}$ people & doc pre-computing & +71ms \\
  \bottomrule
\end{tabular}
\vspace{-1mm}
\end{table}

\begin{table}
\footnotesize
  \caption{\small Offline experiments of DeText-CNN on job recommendation and query auto completion datasets. Both improvements are statistically significant at p < $0.05$.}
\vspace{-2mm}
\label{table:offline-other}
  \begin{tabular}{lll}
    \toprule
     \textbf{Tasks} & \textbf{Metrics} & \textbf{Percentage Lift}\\
    \midrule
    Job Recommendation & AUC & $+3.01\%$ \\
    Query Auto Completion & MRR@10 & $+4.72\%$ \\
  \bottomrule
\end{tabular}
\vspace{-1mm}
\end{table}

\subsection{Extension of DeText to Other Tasks}
In this section, we show the great potential of applying DeText to applications beyond search ranking. We conducted extra experiments on two additional ranking tasks: job recommendation and query auto completion from job search.

For job recommendation, we model the job application probability. The input is a tuple of user id, job post id, and whether the user applied for the job or not. The source fields are from user profiles, including headline, job title, company, and skill. The target fields are from job posts, including job title, job company, job skill, and job country. We used logistic regression as a baseline that is close to production setting, and evaluated with AUC \cite{fawcett2006} metrics. For fair comparison, point-wise ranking (binary classification) is used with no hidden layer of MLP in DeText. Traditional features are kept the same as in the baseline model.


For query auto completion, the source fields are from member profiles, including headline, job title, and company; the target is the completed query. The baseline model is XGBoost with traditional hand-crafted features. We used the same set of traditional features in DeText with listwise LTR, and evaluated with MRR@10 \cite{bar2011}, which is the reciprocal of the rank position of the correct answer.


Table \ref{table:offline-other} shows the offline results. DeText outperforms the baseline models in both tasks by a large margin, indicating that DeText is flexible enough to be applied in other ranking tasks.


\section{Lessons Learned}
We have conducted various experiments on several ranking tasks, where multiple practical methods are used regarding offline relevance, online deployment, latency optimization, etc.  In this section, we summarize the interesting findings and practical solutions into lessons, which could be helpful for both academic and industry practitioners who apply deep NLP models for ranking tasks.

\vspace{1mm}
\noindent\textbf{Deep NLP model performance w.r.t. language genre.} Deep NLP models, especially BERT, are strong at handling paraphrasing. Help center is a good fit, since the queries are close to natural language with rich variation. For people search where queries are mostly named entities, the improvement of both CNN and BERT is smaller. Job search lies in between.

\vspace{1mm}
\noindent\textbf{Pretraining BERT on in-domain data makes a big relevance difference. } 
The common practice of using BERT is to pretrain on general domain such as Wikipedia, and then fine-tune it for a specific task. 
Our experiments suggest that for vertical search systems, it is better to pretrain BERT on in-domain data. Table \ref{table:linkedin-bert} shows that, with only 1/3 of the parameters of BERT$_{\text{BASE}}$, LiBERT significantly outperforms BERT$_{\text{BASE}}$ on people search and job search, while reaching a similar performance on help center.

\vspace{1mm}
\noindent\textbf{Handling traditional features. } Production models are strong and robust with many hand-crafted traditional features.  We observed that 1). after carefully handling these features (Table \ref{table:traditional-features}), deep ranking models can achieve better performance than the production models. 
2). When combining the traditional features with the BERT model, different learning rates should be used.

\vspace{1mm}
\noindent\textbf{Latency reduction solutions. } Latency is one of the biggest challenges to productionize deep learning models, especially the search ranking tasks that involve many documents for one search. In this paper, we present several effective solutions:
\begin{itemize}
    \item For heavy models such as BERT, document pre-computing can save a large amount of computation. Note that the prerequisite is representation based structure.
    \item With two pass ranking, we can deploy a compact CNN based ranking model for real time inference in production for both queries and documents. 
    \item Pretraining a BERT model on in-domain data can maintain the same level of relevance performance, while significantly reducing computation.
\end{itemize}

\section{Conclusions}
In this paper, we propose the DeText (deep text) ranking framework with BERT/CNN based ranking model for practical usage in industry. To accommodate the requirements of different ranking productions, DeText allows flexible configuration, such as input data, text embedding extraction, traditional feature handling, etc. These choices enable us to experiment and develop scalable neural network models with strong relevance performance.  Our offline experiments show that DeText-LiBERT/DeText-CNN consistently outperforms the strong production baselines. The resulting models are deployed into three vertical searches in LinkedIn's commercial search engines. 

%
\bibliographystyle{ACM-Reference-Format}
\bibliography{short}


\begin{thebibliography}{30}


\ifx \showCODEN    \undefined \def \showCODEN     #1{\unskip}     \fi
\ifx \showDOI      \undefined \def \showDOI       #1{#1}\fi
\ifx \showISBNx    \undefined \def \showISBNx     #1{\unskip}     \fi
\ifx \showISBNxiii \undefined \def \showISBNxiii  #1{\unskip}     \fi
\ifx \showISSN     \undefined \def \showISSN      #1{\unskip}     \fi
\ifx \showLCCN     \undefined \def \showLCCN      #1{\unskip}     \fi
\ifx \shownote     \undefined \def \shownote      #1{#1}          \fi
\ifx \showarticletitle \undefined \def \showarticletitle #1{#1}   \fi
\ifx \showURL      \undefined \def \showURL       {\relax}        \fi
\providecommand\bibfield[2]{#2}
\providecommand\bibinfo[2]{#2}
\providecommand\natexlab[1]{#1}
\providecommand\showeprint[2][]{arXiv:#2}

\bibitem[\protect\citeauthoryear{Aksoy and Haralick}{Aksoy and
  Haralick}{2001}]%
        {aksoy2001}
\bibfield{author}{\bibinfo{person}{Selim Aksoy} {and} \bibinfo{person}{Robert~M
  Haralick}.} \bibinfo{year}{2001}\natexlab{}.
\newblock \showarticletitle{Feature normalization and likelihood-based
  similarity measures for image retrieval}.
\newblock \bibinfo{journal}{\emph{Pattern recognition letters}}
  (\bibinfo{year}{2001}).
\newblock


\bibitem[\protect\citeauthoryear{Bar-Yossef and Kraus}{Bar-Yossef and
  Kraus}{2011}]%
        {bar2011}
\bibfield{author}{\bibinfo{person}{Ziv Bar-Yossef} {and} \bibinfo{person}{Naama
  Kraus}.} \bibinfo{year}{2011}\natexlab{}.
\newblock \showarticletitle{Context-sensitive query auto-completion}. In
  \bibinfo{booktitle}{\emph{WWW}}.
\newblock


\bibitem[\protect\citeauthoryear{Burges}{Burges}{2010}]%
        {burges2010}
\bibfield{author}{\bibinfo{person}{Christopher~JC Burges}.}
  \bibinfo{year}{2010}\natexlab{}.
\newblock \showarticletitle{From ranknet to lambdarank to lambdamart: An
  overview}.
\newblock \bibinfo{journal}{\emph{Learning}}  \bibinfo{volume}{11}
  (\bibinfo{year}{2010}).
\newblock


\bibitem[\protect\citeauthoryear{Burges, Ragno, and Le}{Burges
  et~al\mbox{.}}{2007}]%
        {burges2007}
\bibfield{author}{\bibinfo{person}{Christopher~J Burges},
  \bibinfo{person}{Robert Ragno}, {and} \bibinfo{person}{Quoc~V Le}.}
  \bibinfo{year}{2007}\natexlab{}.
\newblock \showarticletitle{Learning to rank with nonsmooth cost functions}. In
  \bibinfo{booktitle}{\emph{NeurIPS}}.
\newblock


\bibitem[\protect\citeauthoryear{Chen and Guestrin}{Chen and Guestrin}{2016}]%
        {chen2016}
\bibfield{author}{\bibinfo{person}{Tianqi Chen} {and} \bibinfo{person}{Carlos
  Guestrin}.} \bibinfo{year}{2016}\natexlab{}.
\newblock \showarticletitle{Xgboost: A scalable tree boosting system}. In
  \bibinfo{booktitle}{\emph{KDD}}.
\newblock


\bibitem[\protect\citeauthoryear{Dai and Callan}{Dai and Callan}{2019}]%
        {dai2019}
\bibfield{author}{\bibinfo{person}{Zhuyun Dai} {and} \bibinfo{person}{Jamie
  Callan}.} \bibinfo{year}{2019}\natexlab{}.
\newblock \showarticletitle{Deeper Text Understanding for IR with Contextual
  Neural Language Modeling}. In \bibinfo{booktitle}{\emph{SIGIR}}.
\newblock


\bibitem[\protect\citeauthoryear{Dai, Xiong, Callan, and Liu}{Dai
  et~al\mbox{.}}{2018}]%
        {Dai:18}
\bibfield{author}{\bibinfo{person}{Zhuyun Dai}, \bibinfo{person}{Chenyan
  Xiong}, \bibinfo{person}{Jamie Callan}, {and} \bibinfo{person}{Zhiyuan Liu}.}
  \bibinfo{year}{2018}\natexlab{}.
\newblock \showarticletitle{Convolutional neural networks for soft-matching
  n-grams in ad-hoc search}. In \bibinfo{booktitle}{\emph{WSDM}}.
\newblock


\bibitem[\protect\citeauthoryear{Devlin, Chang, Lee, and Toutanova}{Devlin
  et~al\mbox{.}}{2019}]%
        {devlin2019}
\bibfield{author}{\bibinfo{person}{Jacob Devlin}, \bibinfo{person}{Ming-Wei
  Chang}, \bibinfo{person}{Kenton Lee}, {and} \bibinfo{person}{Kristina
  Toutanova}.} \bibinfo{year}{2019}\natexlab{}.
\newblock \showarticletitle{BERT: Pre-training of Deep Bidirectional
  Transformers for Language Understanding}. In
  \bibinfo{booktitle}{\emph{NAACL}}.
\newblock


\bibitem[\protect\citeauthoryear{Fawcett}{Fawcett}{2006}]%
        {fawcett2006}
\bibfield{author}{\bibinfo{person}{Tom Fawcett}.}
  \bibinfo{year}{2006}\natexlab{}.
\newblock \showarticletitle{An introduction to ROC analysis}.
\newblock \bibinfo{journal}{\emph{Pattern recognition letters}}
  (\bibinfo{year}{2006}), \bibinfo{pages}{861--874}.
\newblock


\bibitem[\protect\citeauthoryear{Grbovic and Cheng}{Grbovic and Cheng}{2018}]%
        {Grbovic2018}
\bibfield{author}{\bibinfo{person}{Mihajlo Grbovic} {and}
  \bibinfo{person}{Haibin Cheng}.} \bibinfo{year}{2018}\natexlab{}.
\newblock \showarticletitle{Real-time personalization using embeddings for
  search ranking at airbnb}. In \bibinfo{booktitle}{\emph{KDD}}.
\newblock


\bibitem[\protect\citeauthoryear{Guo, Fan, Ai, and Croft}{Guo
  et~al\mbox{.}}{2016}]%
        {Guo:16}
\bibfield{author}{\bibinfo{person}{Jiafeng Guo}, \bibinfo{person}{Yixing Fan},
  \bibinfo{person}{Qingyao Ai}, {and} \bibinfo{person}{W~Bruce Croft}.}
  \bibinfo{year}{2016}\natexlab{}.
\newblock \showarticletitle{A deep relevance matching model for ad-hoc
  retrieval}. In \bibinfo{booktitle}{\emph{CIKM}}.
\newblock


\bibitem[\protect\citeauthoryear{Hochreiter and Schmidhuber}{Hochreiter and
  Schmidhuber}{1997}]%
        {Hochreiter:1997}
\bibfield{author}{\bibinfo{person}{Sepp Hochreiter} {and}
  \bibinfo{person}{Jürgen Schmidhuber}.} \bibinfo{year}{1997}\natexlab{}.
\newblock \showarticletitle{Long short-term memory}. In
  \bibinfo{booktitle}{\emph{Neural computation}}.
\newblock


\bibitem[\protect\citeauthoryear{Huang, He, Gao, Deng, Acero, and Heck}{Huang
  et~al\mbox{.}}{2013}]%
        {Huang:13}
\bibfield{author}{\bibinfo{person}{Po-Sen Huang}, \bibinfo{person}{Xiaodong
  He}, \bibinfo{person}{Jianfeng Gao}, \bibinfo{person}{Li Deng},
  \bibinfo{person}{Alex Acero}, {and} \bibinfo{person}{Larry Heck}.}
  \bibinfo{year}{2013}\natexlab{}.
\newblock \showarticletitle{Learning deep structured semantic models for web
  search using clickthrough data}. In \bibinfo{booktitle}{\emph{CIKM}}.
\newblock


\bibitem[\protect\citeauthoryear{Kim}{Kim}{2014}]%
        {kim2014}
\bibfield{author}{\bibinfo{person}{Yoon Kim}.} \bibinfo{year}{2014}\natexlab{}.
\newblock \showarticletitle{Convolutional Neural Networks for Sentence
  Classification}. In \bibinfo{booktitle}{\emph{EMNLP}}.
\newblock


\bibitem[\protect\citeauthoryear{Kingma and Ba}{Kingma and Ba}{2015}]%
        {kingma2015}
\bibfield{author}{\bibinfo{person}{Diederick~P Kingma} {and}
  \bibinfo{person}{Jimmy Ba}.} \bibinfo{year}{2015}\natexlab{}.
\newblock \showarticletitle{Adam: A method for stochastic optimization}. In
  \bibinfo{booktitle}{\emph{ICLR}}.
\newblock


\bibitem[\protect\citeauthoryear{LeCun and Bengio}{LeCun and Bengio}{1995}]%
        {Lecun:95}
\bibfield{author}{\bibinfo{person}{Yann LeCun} {and} \bibinfo{person}{Yoshua
  Bengio}.} \bibinfo{year}{1995}\natexlab{}.
\newblock \showarticletitle{Convolutional networks for images, speech, and time
  series}.
\newblock \bibinfo{journal}{\emph{The handbook of brain theory and neural
  networks}} (\bibinfo{year}{1995}).
\newblock


\bibitem[\protect\citeauthoryear{Li, Qin, Wang, and Metzler}{Li
  et~al\mbox{.}}{2019}]%
        {li2019}
\bibfield{author}{\bibinfo{person}{Pan Li}, \bibinfo{person}{Zhen Qin},
  \bibinfo{person}{Xuanhui Wang}, {and} \bibinfo{person}{Donald Metzler}.}
  \bibinfo{year}{2019}\natexlab{}.
\newblock \showarticletitle{Combining Decision Trees and Neural Networks for
  Learning-to-Rank in Personal Search}. In \bibinfo{booktitle}{\emph{KDD}}.
\newblock


\bibitem[\protect\citeauthoryear{Nogueira and Cho}{Nogueira and Cho}{2019}]%
        {nogueira2019}
\bibfield{author}{\bibinfo{person}{Rodrigo Nogueira} {and}
  \bibinfo{person}{Kyunghyun Cho}.} \bibinfo{year}{2019}\natexlab{}.
\newblock \showarticletitle{Passage Re-ranking with BERT}.
\newblock \bibinfo{journal}{\emph{arXiv preprint arXiv:1901.04085}}
  (\bibinfo{year}{2019}).
\newblock


\bibitem[\protect\citeauthoryear{Pal and Mitra}{Pal and Mitra}{1992}]%
        {pal1992}
\bibfield{author}{\bibinfo{person}{Sankar~K Pal} {and}
  \bibinfo{person}{Sushmita Mitra}.} \bibinfo{year}{1992}\natexlab{}.
\newblock \showarticletitle{Multilayer perceptron, fuzzy sets, and
  classification}.
\newblock \bibinfo{journal}{\emph{IEEE Transactions on neural networks}}
  (\bibinfo{year}{1992}).
\newblock


\bibitem[\protect\citeauthoryear{Palangi, Deng, Shen, Gao, He, Chen, Song, and
  Ward}{Palangi et~al\mbox{.}}{2016}]%
        {Palangi:16}
\bibfield{author}{\bibinfo{person}{Hamid Palangi}, \bibinfo{person}{Li Deng},
  \bibinfo{person}{Yelong Shen}, \bibinfo{person}{Jianfeng Gao},
  \bibinfo{person}{Xiaodong He}, \bibinfo{person}{Jianshu Chen},
  \bibinfo{person}{Xinying Song}, {and} \bibinfo{person}{Rabab Ward}.}
  \bibinfo{year}{2016}\natexlab{}.
\newblock \showarticletitle{Deep sentence embedding using long short-term
  memory networks: Analysis and application to information retrieval}.
\newblock \bibinfo{journal}{\emph{TASLP}} (\bibinfo{year}{2016}).
\newblock


\bibitem[\protect\citeauthoryear{Pennington, Socher, and Manning}{Pennington
  et~al\mbox{.}}{2014}]%
        {pennington2014}
\bibfield{author}{\bibinfo{person}{Jeffrey Pennington},
  \bibinfo{person}{Richard Socher}, {and} \bibinfo{person}{Christopher~D
  Manning}.} \bibinfo{year}{2014}\natexlab{}.
\newblock \showarticletitle{Glove: Global vectors for word representation}. In
  \bibinfo{booktitle}{\emph{EMNLP}}.
\newblock


\bibitem[\protect\citeauthoryear{Qiao, Xiong, Liu, and Liu}{Qiao
  et~al\mbox{.}}{2019}]%
        {qiao2019}
\bibfield{author}{\bibinfo{person}{Yifan Qiao}, \bibinfo{person}{Chenyan
  Xiong}, \bibinfo{person}{Zhenghao Liu}, {and} \bibinfo{person}{Zhiyuan Liu}.}
  \bibinfo{year}{2019}\natexlab{}.
\newblock \showarticletitle{Understanding the Behaviors of BERT in Ranking}.
\newblock \bibinfo{journal}{\emph{arXiv preprint arXiv:1904.07531}}
  (\bibinfo{year}{2019}).
\newblock


\bibitem[\protect\citeauthoryear{Ramanath, Inan, Polatkan, Hu, Guo, Ozcaglar,
  Wu, Kenthapadi, and Geyik}{Ramanath et~al\mbox{.}}{2018}]%
        {ramanath2018}
\bibfield{author}{\bibinfo{person}{Rohan Ramanath}, \bibinfo{person}{Hakan
  Inan}, \bibinfo{person}{Gungor Polatkan}, \bibinfo{person}{Bo Hu},
  \bibinfo{person}{Qi Guo}, \bibinfo{person}{Cagri Ozcaglar},
  \bibinfo{person}{Xianren Wu}, \bibinfo{person}{Krishnaram Kenthapadi}, {and}
  \bibinfo{person}{Sahin~Cem Geyik}.} \bibinfo{year}{2018}\natexlab{}.
\newblock \showarticletitle{Towards Deep and Representation Learning for Talent
  Search at LinkedIn}. In \bibinfo{booktitle}{\emph{CIKM}}.
\newblock


\bibitem[\protect\citeauthoryear{Sennrich, Haddow, and Birch}{Sennrich
  et~al\mbox{.}}{2016}]%
        {sennrich2016}
\bibfield{author}{\bibinfo{person}{Rico Sennrich}, \bibinfo{person}{Barry
  Haddow}, {and} \bibinfo{person}{Alexandra Birch}.}
  \bibinfo{year}{2016}\natexlab{}.
\newblock \showarticletitle{Neural Machine Translation of Rare Words with
  Subword Units}. In \bibinfo{booktitle}{\emph{ACL}}.
\newblock


\bibitem[\protect\citeauthoryear{Shacham, Merhav, He, and Jiang}{Shacham
  et~al\mbox{.}}{2017}]%
        {shacham2017}
\bibfield{author}{\bibinfo{person}{Dan Shacham}, \bibinfo{person}{Uri Merhav},
  \bibinfo{person}{Qi He}, {and} \bibinfo{person}{Angela Jiang}.}
  \bibinfo{year}{2017}\natexlab{}.
\newblock \bibinfo{title}{Context-aware map from entities to canonical forms}.
\newblock
\newblock
\newblock
\shownote{US Patent App. 15/189,974.}


\bibitem[\protect\citeauthoryear{Shen, He, Gao, Deng, and Mesnil}{Shen
  et~al\mbox{.}}{2014}]%
        {Shen:14}
\bibfield{author}{\bibinfo{person}{Yelong Shen}, \bibinfo{person}{Xiaodong He},
  \bibinfo{person}{Jianfeng Gao}, \bibinfo{person}{Li Deng}, {and}
  \bibinfo{person}{Gr{\'e}goire Mesnil}.} \bibinfo{year}{2014}\natexlab{}.
\newblock \showarticletitle{A latent semantic model with convolutional-pooling
  structure for information retrieval}. In \bibinfo{booktitle}{\emph{CIKM}}.
\newblock


\bibitem[\protect\citeauthoryear{Vaswani, Shazeer, Parmar, Uszkoreit, Jones,
  Gomez, Kaiser, and Polosukhin}{Vaswani et~al\mbox{.}}{2017}]%
        {vaswani:17}
\bibfield{author}{\bibinfo{person}{Ashish Vaswani}, \bibinfo{person}{Noam
  Shazeer}, \bibinfo{person}{Niki Parmar}, \bibinfo{person}{Jakob Uszkoreit},
  \bibinfo{person}{Llion Jones}, \bibinfo{person}{Aidan~N Gomez},
  \bibinfo{person}{{\L}ukasz Kaiser}, {and} \bibinfo{person}{Illia
  Polosukhin}.} \bibinfo{year}{2017}\natexlab{}.
\newblock \showarticletitle{Attention is all you need}. In
  \bibinfo{booktitle}{\emph{NeurIPS}}.
\newblock


\bibitem[\protect\citeauthoryear{Xiong, Dai, Callan, Liu, and Power}{Xiong
  et~al\mbox{.}}{2017}]%
        {xiong2017}
\bibfield{author}{\bibinfo{person}{Chenyan Xiong}, \bibinfo{person}{Zhuyun
  Dai}, \bibinfo{person}{Jamie Callan}, \bibinfo{person}{Zhiyuan Liu}, {and}
  \bibinfo{person}{Russell Power}.} \bibinfo{year}{2017}\natexlab{}.
\newblock \showarticletitle{End-to-end neural ad-hoc ranking with kernel
  pooling}. In \bibinfo{booktitle}{\emph{SIGIR}}.
\newblock


\bibitem[\protect\citeauthoryear{Yin, Hu, Tang, Daly, Zhou, Ouyang, Chen, Kang,
  Deng, Nobata, et~al\mbox{.}}{Yin et~al\mbox{.}}{2016}]%
        {yin2016}
\bibfield{author}{\bibinfo{person}{Dawei Yin}, \bibinfo{person}{Yuening Hu},
  \bibinfo{person}{Jiliang Tang}, \bibinfo{person}{Tim Daly},
  \bibinfo{person}{Mianwei Zhou}, \bibinfo{person}{Hua Ouyang},
  \bibinfo{person}{Jianhui Chen}, \bibinfo{person}{Changsung Kang},
  \bibinfo{person}{Hongbo Deng}, \bibinfo{person}{Chikashi Nobata},
  {et~al\mbox{.}}} \bibinfo{year}{2016}\natexlab{}.
\newblock \showarticletitle{Ranking relevance in yahoo search}. In
  \bibinfo{booktitle}{\emph{KDD}}.
\newblock


\bibitem[\protect\citeauthoryear{Zamani, Mitra, Song, Craswell, and
  Tiwary}{Zamani et~al\mbox{.}}{2018}]%
        {Zamani2018}
\bibfield{author}{\bibinfo{person}{Hamed Zamani}, \bibinfo{person}{Bhaskar
  Mitra}, \bibinfo{person}{Xia Song}, \bibinfo{person}{Nick Craswell}, {and}
  \bibinfo{person}{Saurabh Tiwary}.} \bibinfo{year}{2018}\natexlab{}.
\newblock \showarticletitle{Neural ranking models with multiple document
  fields}. In \bibinfo{booktitle}{\emph{WSDM}}.
\newblock


\end{thebibliography}

\end{document}